\documentclass[aps,prb,twocolumn,showpacs,superscriptaddress]{revtex4}
\usepackage{epsfig}
\usepackage{graphicx}
\usepackage{amsfonts}
\usepackage[figuresright]{rotating}
\usepackage{amssymb}
\usepackage{amsmath}

\usepackage{psfrag}
\newtheorem{theorem}{Theorem}
\newtheorem{lemma}{Lemma}
\newtheorem{corollary}{Corollary}
\newenvironment{remark}[1][Remark]{\noindent \textbf{#1: }}{\ \rule{0.5em}{0.5em}}
\newenvironment{proof}[1][Proof]{\noindent \textbf{#1: }}{\ \rule{0.5em}{0.5em}}

\def\avg#1{\langle#1\rangle}

\def\be{\begin{equation}} \def\ee{\end{equation}}
\def\bea{\begin{eqnarray}} \def\eea{\end{eqnarray}}

\def\nn{\nonumber}
\def\pp{\parallel}

\newcommand{\ket}[1]{| #1 \rangle}

\begin{document}
\title{Exact Results on Itinerant Ferromagnetism in 
Multi-orbital Systems on Square and Cubic Lattices}
\author{Yi Li}
\affiliation{Princeton Center for Theoretical Science, Princeton University, Princeton, NJ 08544}
\affiliation{Department of Physics, University of California, San Diego,
La Jolla, CA 92093}
\author{Elliott H. Lieb}
\affiliation{Departments of Mathematics and Physics, Princeton University, Princeton, NJ 08544}
\author{Congjun Wu}
\affiliation{Department of Physics, University of California, San Diego,
La Jolla, CA 92093}
\date{April 30, 2014}

\begin{abstract}
We study itinerant ferromagnetism in multi-orbital Hubbard models in
certain two-dimensional square and three-dimensional cubic
lattices. 
In the strong coupling limit where doubly occupied orbitals are not allowed,
we prove that the fully spin-polarized states are the unique ground states, apart from the trivial spin degeneracies, for a large region of fillings factors.
Possible applications to $p$-orbital bands with ultra-cold fermions in optical
lattices, and electronic $3d$-orbital bands in transition-metal oxides,
are discussed.
\end{abstract}

\maketitle

Itinerant ferromagnetism (FM) is one of the central topics in condensed
matter physics \cite{lieb1962,mattis2006,nagaoka1966,roth1966,kugel1973,
hertz1976,moriya1985,torrance1987,gill1987,hirsch1989,shastry1990, mielke1991,mielke1991a,tasaki1992, millis1993,belitz2005,lohneysen2007,liu2012}.
Historically, it had been thought that exchange energy, which is a
perturbation-theoretic idea,
favors FM, but that is opposed by the kinetic energy increase
required by the Pauli exclusion principle to polarize a
fermionic system.
Interactions need to be sufficiently strong to drive FM transitions,
and hence FM is intrinsically a strong correlation problem.
In fact, the Lieb-Mattis theorem \cite{lieb1962} for one-dimensional (1D)
systems shows that FM never occurs, regardless of how large the exchange energy
might be.
Even with very strong repulsion, electrons can remain unpolarized
while their wave functions are nevertheless significantly far from the
Slater-determinant type.

Strong correlations are necessary for itinerant FM but the precise mechanism
is subtle. 
An early example 
is Nagaoka's theorem about the infinite $U$ Hubbard model, fully filled except
for one missing electron, called a hole.
He showed \cite{nagaoka1966}, and Tasaki generalized the result \cite{Tasaki1989}, that
the one hole causes the system of itinerant electrons to be fully
spin-polarized - \textit{i.e.}, saturated FM.
However, Nagaoka's theorem is not relevant in 1D, 
because
no non-trivial loops are possible in this case.
For infinite $U$, ground states are degenerate regardless of spin
configurations along the chain.
As $U$ becomes finite, as shown in Ref. \onlinecite{lieb1962a},
the degeneracy is lifted and the ground state is a spin singlet.
Another set of exact results are the flat-band FM models on line graphs
\cite{mielke1991,mielke1991a,mielke1992,mielke1993,tasaki1992}.
On such graphs, there exist Wannier-like localized single particle eigenstates,
which eliminate the kinetic energy cost of spin polarization.
Later, interesting metallic ferromagnetic models without flat band
structures have been
proposed by Tasaki\cite{Tasaki2003}, Tanaka and Tasaki\cite{Tanaka2007}. 
FM in realistic flat-band systems has been proposed
in the $p$-orbitals in honeycomb lattices with ultra-cold
fermions \cite{zhangSZ2010}.

In this article, we prove a theorem about FM in the two-dimensional (2D) square
and three-dimensional (3D) cubic lattices with \textit{multi-orbital}
structures.
We can even do this in 1D, as shows in Corollary 2
in Appendix \ref{sect:1dFM},
where we reproduce, by our method, Shen's result \cite{shen1998}
that the \textit{multi-orbital} 1D system is FM.
Our result differs from Nagaoka's in that it is valid for
a large region of filling factors in both 2D and 3D.
It is also different from flat band FM, in which fermion
kinetic energy differences are suppressed.

We emphasize that our result is robust in that the translation invariance
is not really required.
The hopping magnitudes can vary along chains and from chain to chain.
We confine our attention here to translation invariant Hamiltonian purely for simplicity of exposition.

Our band Hamiltonians behave like decoupled, perpendicular 1D chains,
which are coupled by the standard on-site, \textit{multi-orbital}
Hubbard interactions that are widely used in the literature
\cite{roth1966,kugel1973,cyrot1975,oles1983}.
In the limit of infinite intra-orbital repulsion, we prove that the
inter-orbital Hund's rule coupling at each site drives the ground
states to fully spin-polarized states.
Furthermore, the ground states are non-degenerate except for the obvious
spin degeneracy, and the wave functions are nodeless in a properly defined basis.
This theorem is generalized here to multi-component fermions with
SU($N$) symmetries.
This itinerant FM theorem is not just of academic interest
because it may be relevant to the $p$-orbital systems with ultra-cold atoms\cite{wang2008}
and to the LaAlO$_3$/SrTiO$_3$ interface of $3d$-orbital transition-metal oxides\cite{lilu2011,bert2011,chen2013}.



Let us first very briefly give a heuristic
overview of our model in 2D.
Think of the square lattice $\mathbb{Z}^2$ as consisting of horizontal lines
and vertical lines, and imagine two kinds of electrons, one of which can
move with hard-core interactions along the horizontal lines, and the other
of which can move along the vertical lines.
No transition between any two lines is allowed. When two electrons of different
type meet at a vertex, Hund's rule requires them to prefer to be in a triplet
state.
Our theorem is that this interaction forces the whole system to be uniquely FM.
The two kinds of electrons in this picture are the $p_x$-orbital and
$p_y$-orbital electrons.
The $p_x$ orbitals overlap only in the $x$-direction and thus can allow motion
only in that direction - and similarly for $p_y$ orbitals.

\begin{figure}[tbp]
\centering\epsfig{file=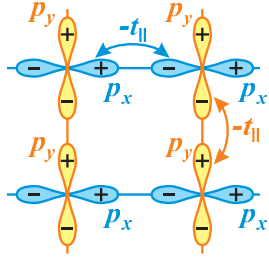,clip=1,width=0.4\linewidth,angle=0}
\caption{The square lattice with the quasi-1D band structure of the $p$-orbital bands.
Particles in the $p_x (p_y)$-orbital can only move along the
$x (y)$-direction, respectively.
The sign of $t_{\pp}$ can be changed by a gauge transformation on
the square lattice.
}
\label{fig:orbital}
\end{figure}

Now, let us describe \textit{multi-orbital} systems for spin-$\frac{1}{2}$ fermions on 2D
square and 3D cubic lattices with quasi-1D band structures.
The $p$-orbital systems are used, but this is only one
possible example of atomic orbitals that could be considered,
another example being $d_{xz}$- and $d_{yz}$- orbitals.
Nearest-neighbor (NN) hoppings can be classified as either
$\sigma$-bonding with hopping amplitude $t_\pp$ or
$\pi$-bonding with hopping amplitude $t_\perp$, which describe
the hopping directions parallel or perpendicular to the orbital
orientation, respectively.
Typically, $t_\perp$ is much smaller than $t_\pp$, and thus will be
neglected here, leading to the following quasi-1D band Hamiltonian
(see Fig. \ref{fig:orbital}):
\bea
H_{kin}^{2D(3D)}&=& \sum_{\mu=x,y,(z)} H_{kin}^{1D,\mu}
-\mu_0\sum_{\mathbf{r}} n(\mathbf{r}), \ \ \, \ \ \, \nn \\
H_{kin}^{1D,\mu}
&=&-t_\pp
\sum_{\mathbf{r}, \sigma=\uparrow,\downarrow} p_{\mu,\sigma}^\dagger(\mathbf{r}
+\hat{\mathbf{e}}_{\mu})p_{\mu,\sigma}(\mathbf{r})+ h.c.. \ \ \,
\label{eq:H_band}
\eea
Here, $p_{\mu,\sigma}(\mathbf{r})$ is the annihilation operator in the
$p_{\mu}$-orbital ($\mu=x,y,(z)$) on site $\mathbf{r}$
with the spin eigenvalue $\sigma$;
$n(\mathbf{r})$ is the total particle number on site $\mathbf{r}$,
$\hat{\mathbf{e}}_{\mu}$ is the unit vector in the $\mu$-direction.
Since the lattice is bipartite, the sign of $t_\pp$ can be flipped
by a gauge transformation.
Without loss of generality, it is taken to be positive.
The generic \textit{multi-orbital} on-site Hubbard interactions \cite{hubbard1963,slater1963} are as follows:
\bea
H_{int}&=&
U \sum_{\mu, \mathbf{r}}
n_{\mu,\uparrow}(\mathbf{r}) n_{\mu,\downarrow}(\mathbf{r})
+ \frac{V}{2}\sum_{\mu \neq \nu, \mathbf{r}} n_{\mu} (\mathbf{r})
n_{\nu}(\mathbf{r})\nn \\
&-& \frac{J}{2} \sum_{\mu \neq \nu, \mathbf{r}} \Big\{ \vec S_{\mu}
(\mathbf{r}) \cdot \vec S_{\nu}(\mathbf{r}) -
\frac{1}{4} n_{\mu}(\mathbf{r}) n_{\nu}(\mathbf{r})\Big\} \nn \\
&+& \Delta \sum_{\mu \neq \nu, \mathbf{r}}
p^\dagger_{\mu \uparrow} (\mathbf{r}) p^\dagger_{\mu \downarrow}(\mathbf{r})
p_{\nu \downarrow}(\mathbf{r}) p_{\nu \uparrow}(\mathbf{r})
,
\label{eq:H_int}
\eea
where $n_{\mu,\sigma}=p_{\mu,\sigma}^{\dagger} p_{\mu,\sigma}$;
$\vec S_{\mu}=p_{\mu,\alpha}^{\dagger}\vec S_{\alpha \beta}
p_{\mu,\beta}$ represents
the spin operators in the $p_{\mu}$-orbital.
The $U$ and $V$ terms are intra- and inter-orbital Hubbard interactions,
respectively;
the $J$ term represents Hund's rule coupling;
the $\Delta$ term describes the pair hopping process between different orbitals.
The expressions of $U$, $V$, $J$, and $\Delta$ in terms of integrals of
Wannier orbital wave functions and their physical meaning
are provided in Appendix \ref{sect:int_para}.


We consider the limit  $U \rightarrow +\infty$ and
start with the 2D version of the Hamiltonian $H_{kin}+H_{int}$.
States with double occupancy in a \textit{single} orbital,
$\frac{1}{\sqrt 2} \left \{p_{x\uparrow}^\dagger p_{x\downarrow}^\dagger\pm
p_{y\uparrow}^\dagger p_{y\downarrow}^\dagger \right\} \ket{0}$, are projected out.
The projected Fock space on a single site is a tensor product of
that on each orbital spanned by three states
as ${\cal F}_{\mathbf{r}}=\bigotimes_{\mu=x,y}{\cal F}_{\mathbf{r}}^{\mu}$
with ${\cal F}_{\mathbf{r}}^{\mu}=\{\ket{0},
p_{\mu, \uparrow}^{\dagger} (\mathbf{r})\ket{0},
p_{\mu, \downarrow}^{\dagger} (\mathbf{r})\ket{0} \}$.
The projected Fock space $\cal F$ of the system is a
tensor product of ${\cal F}_{\mathbf{r}}$ on each site.

We state three lemmas before presenting the FM Theorem 1.
The proofs of Lemmas 2 and 3 are provided in
Appendix \ref{app:lemmas}.
We shall always assume henceforth the following \textit{two conditions}
which are essential for Lemma 2 and Lemma 3 below respectively:

($\ast$) \textit{The boundary condition
\footnote{This condition is taken to simplify the proof.
Our results also hold for open boundary conditions without
the constraint on odd/even particle numbers in each row and column.
With open boundary conditions,
Theorem 1 remains correct when both the on-site potentials and NN
hoppings are disordered.
}
on each row and column is periodic (resp. anti-periodic)
when the particle number in the row or column is odd (resp. even).}
The fact that the particle number in each row/column is fixed
is contained in Lemma 1 below.

($\ast \ast$) \textit{There is at least one particle and one hole in each
chain. `Hole' means an empty orbital.}

The following lemma is obvious.

\vspace{-2mm}

{\lemma In the projected Fock space ${\cal F}$ for the Hamiltonian
$H=H_{kin}+H_{int}$ (see Eqs. (\ref{eq:H_band}) and (\ref{eq:H_int})),
the particle numbers of each row and each column are separately conserved.}

Based on Lemma 1, we can specify a partition of particle numbers into
rows $\mathcal{X}=\left\{ r_i =1, \cdots, L_y\right \}$
and columns $\mathcal{Y}=\left\{ c_i =1, \cdots, L_x \right \}$ as,
\bea
\mathcal{N}_{\mathcal{X}}=\left\{ N_{r_i} \right \},
\mathcal{N}_{\mathcal{Y}}=\left\{ N_{c_i} \right \},
\label{eq:partition}
\eea
where $N_{r_i}$ and $N_{c_i}$ are the particle numbers conserved in the
$r_i$-th row and the $c_i$-th column, respectively.
Altogether $\sum_{r_i=1}^{L_y} N_{r_i}+\sum_{c_i=1}^{L_x} N_{c_i}=N_{tot}$,
is the total particle number.
The physical Hilbert space
${\mathcal H}_{\mathcal{N}_{\mathcal{X}}, \mathcal{N}_{\mathcal{Y}}}$ is spanned
by states in $\cal F$ satisfying
Eq. (\ref{eq:partition}).
A many-body basis in ${\cal H}_{\mathcal{N}_{\mathcal{X}},
\mathcal{N}_{\mathcal{Y}}}$ can be defined using the following
convention:
we first order $p_x$-orbital particles in each row
by successively applying creation operators of $p_x$-orbitals,
starting with the left most occupied site $x^r_1$ and continuing to the right until $x^r_{N_r}$
in the $r$-th row.
The operator creating the whole collection of $N_r$ $p_x$-orbital particles in the row $r$ is
denoted as
\bea
&&\mathcal{P}_{x,r}^{\dagger}=\prod_{i=1,\mathbf{r}_i \in \mathrm{row}\, r
}^{N^r}
p_{x, \alpha^r_i}^{\dagger}(\mathbf{r}_i) \nn \\
&=&p_{x, \alpha^r_{N_r}}^{\dagger}(\mathbf{r}_{N_r})
\cdots p_{x, \alpha^r_2}^{\dagger}(\mathbf{r}_2)
p_{x, \alpha^r_1}^{\dagger}(\mathbf{r}_1).
\eea
Here, $i$ is the particle index in row $r$. $\mathbf{r}_i=(x^r_i,r)$ and
$\alpha^r_i$ are respectively the coordinate and  $s_z$
eigenvalue for the $i$-th particle in the $r$-th row;
similarly, the creation operator for the $N^c$ $p_y$-orbital particles in
the $c$-th column can be defined, following an order from top to bottom, as
$\mathcal{P}_{y,c}^{\dagger}=\prod_{i=1,\mathbf{r}_i \in\mathrm{column}\, c}^{N^c}
p_{y, \beta^c_i}^{\dagger}(\mathbf{r}_i)
=p_{y, \beta^c_{N_c}}^{\dagger}(\mathbf{r}_{N_c})
\cdots p_{y, \beta^c_2}^{\dagger}(\mathbf{r}_2)
p_{y, \beta^c_1}^{\dagger}(\mathbf{r}_1).
$
Here, similar definitions apply to $\mathbf{r}_i=(c, y^c_i)$ in column $c$
and $\beta^c_i$.
These coordinates for particles in each chain are ordered as
$1\le x^r_1< x^r_2 < ... < x^r_{N_r}\le L_x$, and
$1\le y^c_1< y^c_2 < ... < y^c_{N_c}\le L_y$.

Based on the above ordering within each row and each column,
the many-body basis can be set up by further ordering
them by rows and columns and applying the following
creation operators to the vacuum $\ket{0}$
as:
\bea
&& 
\ket{\mathcal{R}, \mathcal{S}}
_{\mathcal{N}_{\mathcal{X}}, \mathcal{N}_{\mathcal{Y}}}
=\prod_{j=1}^{L_x} \mathcal{P}_{y,c_j}^{\dagger}
\prod_{j=1}^{L_y} \mathcal{P}_{x,r_j}^{\dagger} \ket{0} \nn \\
&=& \mathcal{P}_{y,c_{L_x}}^{\dagger} \cdots \mathcal{P}_{y,c_2}^{\dagger}
\mathcal{P}_{y,c_1}^{\dagger}
 \mathcal{P}_{x,r_{L_y}}^{\dagger} \cdots \mathcal{P}_{x,r_2}^{\dagger}
 \mathcal{P}_{x,r_1}^{\dagger} \ket{0},
\label{eq:basis}
\eea
Here, $j$ denotes the index of columns and rows. Given a partition of the particle
number $\mathcal{N}_{\mathcal{X}}, \mathcal{N}_{\mathcal{Y}}$,
the many-body basis is specified by the coordinates of occupied sites
$\mathcal{R}=\{ \mathbf{r}^{r_j}_i; \mathbf{r}^{c_j}_i\}$ and the corresponding
spin configuration $\mathcal{S}=\{ \alpha^{r_j}_i; \beta^{c_j}_i\}$ for
all  $i$'s and $j$'s.

\vspace{-2mm}
{\lemma \textbf{(Non-positivity)} The off-diagonal matrix elements of the
Hamiltonian $H_{kin}+H_{int}$ with respect to the bases
defined in Eq. (\ref{eq:basis}) are non-positive.}

Since the Hamiltonian is spin invariant, its eigenstates
can be labeled by the total spin $S$ and its $z$-component $S_z$.
The Hilbert space ${\cal H}_{\mathcal{N}_{\mathcal{X}}, \mathcal{N}_{\mathcal{Y}}}$
can be divided into subspaces with different values of total $S_z$, denoted
as ${\cal H}^{S_z}_{\mathcal{N}_{\mathcal{X}}, \mathcal{N}_{\mathcal{Y}}}$.
The many-body basis in this subspace is denoted as
$\ket{\mathcal{R}, \mathcal{S}}^{S_z}$.
The smallest non-negative value of $S_z$ is denoted as $S_{z}^{min}$,
which equals $0$ ($\frac{1}{2}$) for even (odd) values of $N_{tot}$.
The corresponding subspace is denoted as
${\cal H}^{min}_{\mathcal{N}_{\mathcal{X}}, \mathcal{N}_{\mathcal{Y}}}$.
Every set of eigenstates with total spin $S$ has one representative
in  ${\cal H}^{min}_{\mathcal{N}_{\mathcal{X}}, \mathcal{N}_{\mathcal{Y}}}$, and thus
the ground states in this subspace are also the ground states in the
entire Hilbert space.

\vspace{-2mm}
{\lemma\textbf{(Transitivity)} Consider the Hamiltonian matrix in
the subspace ${\cal H}^M_{\mathcal{N}_{\mathcal{X}}, \mathcal{N}_{\mathcal{Y}}}$
with $S_z=M$.
Under  \textit{condition} ($\ast \ast$), for any two basis vectors
$\ket{u}$
and $\ket{u^\prime}$ 
there exits a series of basis vectors with nonzero matrix elements
$\ket{u_1}, \ket{u_2}, ..., \ket{u_k}$ connecting them, i.e.,
\bea
\avg{u|H|u_1}\avg{u_1|H|u_2}... \avg{u_k|H|u^\prime} \neq 0.
\eea
}
Based on the above lemmas, we now establish the following theorem
about FM, which is the main result of this article.

\vspace{-2mm}
{\theorem\textbf{(2D FM Ground State)} Consider the Hamiltonian $H_{kin}+H_{int}$ with
boundary condition ($\ast$) in the limit $U \rightarrow +\infty$.
The physical Hilbert space is ${\cal H}_{\mathcal{N}_{\mathcal{X}},
\mathcal{N}_{\mathcal{Y}}}$.
For any value of $J>0$, the ground states include the fully spin-polarized
states.
If  \textit{condition} ($\ast \ast$) is also satisfied, the ground state is unique apart
from the trivial spin degeneracy.
The ground state $\ket{\Psi_G^M}$ in
${\cal H}^M_{\mathcal{N}_{\mathcal{X}}, \mathcal{N}_{\mathcal{Y}}}$ for
all 
values of $-N_{tot}/2\le M \le N_{tot}/2$
form a set of spin multiplet with $S=N_{tot}/2$, which
can be expressed as
\bea
\ket{\Psi_G^M}=\sum_{\mathcal{R}, \mathcal{S}}
c_{\mathcal{R}, \mathcal{S}}
\ket{\mathcal{R}, \mathcal{S}}^M
\label{eq:groundstate}
\eea
with all the coefficients strictly positive.
}

{\proof}  Lemma 2 together with the Perron-Frobenius
(PF) theorem \cite{perron1907,frobenius1908} (see Appendix \ref{app:p-f}) implies that
there is a ground state
$\ket{\Psi_G^M}$ in ${\cal H}^{M}_{\mathcal{N}_{\mathcal{X}},
\mathcal{N}_{\mathcal{Y}}}$ that
can be expanded as
\bea
\ket{\Psi_G^{M}}=\sum_{\mathcal{R}, \mathcal{S}}
c_{\mathcal{R}, \mathcal{S}}
\ket{\mathcal{R}, \mathcal{S}}^{M},
\eea
with all coefficients non-negative, {\it i.e.},
$c_{\mathcal{R}, \mathcal{S}}\ge 0$.
Because of the possible degeneracy, $\ket {\Psi_G^{M}}$
may not be an eigenstate of total spin.
We define a reference state by summing over all the bases in
${\cal H}^{M}_{\mathcal{N}_{\mathcal{X}}, \mathcal{N}_{\mathcal{Y}}}$
with equal weight as
$\ket{\Psi_{FM}^M}=\sum_{\mathcal{R}, \mathcal{S}}
 \ket{\mathcal{R}, \mathcal{S}}^{M}$,
which is symmetric under exchange of spin configurations of any two
particles, and thus is one of the multiplet of the
fully polarized states $S=\frac{N_{tot}}{2}$.
Define a projection operator $P_{S}$ for the subspace
spanned by states with total spin $S$.
Clearly, $\avg{\Psi_G^{M}|\Psi_{FM}^M}=\sum_{\mathcal{R}, \mathcal{S}}
c_{\mathcal{R}, \mathcal{S}}>0$ up
to normalization factors, thus $P_{\frac{N_{tot}}{2}} \ket{\Psi_G^M}\neq 0$.
We have
\bea
H  P_{\frac{N_{tot}}{2}} \ket{\Psi_G^M}=  P_{\frac{N_{tot}}{2}} H\ket{\Psi_G^M}
=E_G^M P_{\frac{N_{tot}}{2}} \ket{\Psi_G^M}.
\eea
For $M=S_z^{min}$, $P_{S=\frac{N_{tot}}{2}} \ket{\Psi_G^M}$
lies in  ${\cal H}^{min}_{\mathcal{N}_{\mathcal{X}},
\mathcal{N}_{\mathcal{Y}}}$, and thus is a ground
state in the entire Hilbert space.

Further, if  \textit{condition} ($\ast \ast$) is satisfied, Lemma 3 of
transitivity is also valid. In that case, the Hamiltonian matrix in the
subspace ${\cal H}^M_{\mathcal{N}_{\mathcal{X}}, \mathcal{N}_{\mathcal{Y}}}$
is irreducible.
According to PF
theorem, the ground state $\ket{\Psi_G^M}$
in this subspace is non-degenerate and thus it must be an eigenstate
of total spin which should be $S=N_{tot}/2$.
Otherwise, $\avg{\Psi^{M}_G|\Psi^{M}_{FM}}=0$
which would contradict to the fact that $\avg{\Psi^{M}_G|\Psi^{M}_{FM}}>0$.
Furthermore, the coefficients in the expansion of Eq. (\ref{eq:groundstate})
are strictly positive, {\it i.e.} $c_{\mathcal{R}, \mathcal{S}}> 0$,
as explained in Appendix \ref{app:p-f}.
\hfill \textit{Q.E.D.}

{\remark}
{Theorem 1 does not require translation symmetry and
thus remains true in the presence of on-site disorders.}

Theorem 1 is a joint effect of the 1D band structure and the \textit{multi-orbital}
Hund's rule (\textit{i.e.} $J>0$).
In the usual 1D case, if $U$ is infinite, fermions cannot pass each other.
With periodic boundary conditions, only order-preserving cyclic
permutations of spins can be realized through hopping terms, and thus
the Hamiltonian matrix is \textit{not transitive}.
The ground states are degenerate.
For $H_{kin}+H_{int}$, particles in orthogonal chains meet each other at the
crossing sites, and their spins are encouraged to align by the $J$ term, which
also promotes the transitivity of the Hamiltonian matrix.
This removes the degeneracy and selects the fully polarized
FM state.
If  \textit{condition} ($\ast \ast$) is not met, Lemma 3 of transitivity
may not be valid,
and thus the ground states could be degenerate.
On the other hand,  \textit{condition} ($\ast \ast$) is not necessary for
transitivity,
and can be relaxed to a weaker condition as described in Appendix \ref{app:transitivy}.

Unlike Nagaoka's FM state,
the particles in our FM states still interact with each other through the $V$ term
even though they are fully polarized.
Conceivably, it could further lead to Cooper pairing instability and
other strong correlation phases within the
fully polarized states.
Owing to the nodeless structure of the ground state wavefunction, Eq.
(\ref{eq:groundstate}), these states can be simulated by
quantum Monte Carlo simulations free of any sign problem.


Theorem 1 can be further generalized from the SU(2) systems to those
with SU($N$) symmetry.
These high-spin symmetries are not just of academic interest.
It is proposed to use ultra-cold alkali and alkaline-earth fermions to
realize SU($N$) and Sp($N$) symmetric systems \cite{wu2003,wu2006,
gorshkov2010,wu2012}.
Recently, the SU(6) symmetric $^{173}$Yb fermions have been loaded
into optical lattices to form a Mott-insulating state \cite{taie2010,taie2012}.
The SU($N$) kinetic energy $H_{kin}^{SU}$ can be obtained by simply
increasing the number of fermion components in $H^{1D,\mu}_{kin}$ defined
in Eq. (\ref{eq:H_band}), {\it i.e.}, $\sigma=1, 2, ..., N$.
The SU($N$) interaction term can be expressed as
\bea
H_{int}^{SU}&=& \frac{U}{2} \sum_{\mu, \sigma \neq \sigma^\prime, \mathbf{r}}
n_{\mu,\sigma}(\mathbf{r}) n_{\mu,\sigma^\prime}(\mathbf{r})
+ \frac{V}{2}\sum_{\mu \neq \nu, \mathbf{r}} n_{\mu} (\mathbf{r}) n_{\nu}(\mathbf{r}) \nn \\
&-& \frac{J}{4} \sum_{\mu \neq \nu, \mathbf{r}}
\Big\{P_{\mu\nu} (\mathbf{r})
-n_{\mu}(\mathbf{r}) n_{\nu}(\mathbf{r})\Big\} \nn \\
&+& \frac{\Delta}{2} \sum_{\mu \neq \nu, \sigma \neq \sigma^\prime, \mathbf{r}}
p^\dagger_{\mu \sigma} (\mathbf{r}) p^\dagger_{\mu \sigma^\prime}(\mathbf{r})
p_{\nu \sigma^\prime}(\mathbf{r}) p_{\nu \sigma}(\mathbf{r}), 
\label{eq:H_suint}
\eea
where $n_\mu(\mathbf{r})=\sum_{\sigma} n_{\mu,\sigma}(\mathbf{r})$;
$P_{\mu \nu}(\mathbf{r})$ is the exchange operator defined as
$P_{\mu \nu}(\mathbf{r})=
\sum_{\sigma\sigma^\prime}p^\dagger_{\mu\sigma} (\mathbf{r}) p^\dagger_{\nu\sigma^\prime}
(\mathbf{r}) p_{\mu\sigma^\prime}(\mathbf{r}) p_{\nu\sigma} (\mathbf{r})$.

For the SU($N$) Hamiltonian $H^{SU}_{kin}+H^{SU}_{int}$, not only is the
particle number of each chain separately conserved, but also
the total particle number of each component $\sigma$ is
separately conserved.
We still use $\mathcal{N}_{\mathcal{X}}$ and $\mathcal{N}_{\mathcal{Y}}$
to denote particle number distribution in rows and columns, and use
$\mathcal{N}_{\sigma}$ to represent
the distribution of particle number among different components.
The corresponding subspace is denoted as
${\cal H}_{\mathcal{N}_{\mathcal{X}}, \mathcal{N}_{\mathcal{Y}}}^{\mathcal{N}_{\sigma}}$.
By imitating the proof of Theorem 1, we arrive
at the following theorem.
The proof is shown in Appendix \ref{sect:SUN}.

\vspace{-3mm}
{\theorem \textbf{(SU(N) Ground State FM})
Consider the SU($N$) Hamiltonian $H_{kin}^{SU}+H_{int}^{SU}$
in the limit  $U \rightarrow \infty$, whose
physical Hilbert space is ${\cal H}_{\mathcal{N}_{\mathcal{X}},
\mathcal{N}_{\mathcal{Y}}}$.
Under \textit{condition} ($\ast$), for any value of $J>0$, the
ground states include those belonging
to the fully symmetric rank-$N_{tot}$ tensor representation.
If  \textit{condition} ($\ast \ast$) is further satisfied,
the ground states are unique apart from the trivial
$\frac{(N+N_{tot}-1)!}{(N-1)!N_{tot}!}$-fold SU($N$) spin degeneracy.
In each subspace ${\cal H}_{\mathcal{N}_{\mathcal{X}},
\mathcal{N}_{\mathcal{Y}}}^{\mathcal{N}_{\sigma}}$,
$\ket{\Psi_G^{\mathcal{N}_{\sigma}}}=\sum_{u} c_u \ket{u}$,
with $c_u>0$ for all basis vectors of $\ket{u}$ in the subspace
${\cal H}_{\mathcal{N}_{\mathcal{X}}, \mathcal{N}_{\mathcal{Y}}}^{\mathcal{N}_{\sigma}}$.
}

\vspace{2mm}
We turn now to the 3D and 1D cases.
As proved in Appendix \ref{sect:cubic},
Lemmas 1, 2, and 3 are still valid under  \textit{conditions}
($\ast$) and ($\ast \ast$).
We then arrive at the following corollary.
(The 1D case is discussed in Appendix \ref{sect:1dFM}).
\vspace{-2mm}
{\corollary \textbf{(3D FM Ground State)} The statements in Theorems 1 and 2 of FM are also valid for
the 3D version of $H_{kin}+H_{int}$
defined in Eq. (\ref{eq:H_band}) and Eq. (\ref{eq:H_int})
under the same conditions.
}

So far, we have considered the case of $J>0$.
In certain systems with strong electron-phonon coupling, such as
alkali-doped fullerenes, Hund's rule may be replaced by an
anti-Hund's rule, i.e., $J<0$ \cite{gunnarsson2004}.
In this case, we obtain the following Theorem 3 in 2D.

\vspace{-3mm}
{\theorem
Consider the 2D Hamiltonian $H_{kin} + H_{int}$ in the limit
$U \rightarrow +\infty$ with $J<0$.
If \textit{conditions} ($\ast$) and ($\ast \ast$) are satisfied,
then the ground state
in each subspace ${\cal H}_{\mathcal{N}_{\mathcal{X}},
\mathcal{N}_{\mathcal{Y}}}^{M}$,
denoted as $\ket{\Psi_G^M}$, is non-degenerate
and obeys the following sign rule
\bea
\ket{\Psi_G^M}
&=& \sum_{\mathcal{R}, \mathcal{S}}  (-)^\Gamma
c_{\mathcal{R}, \mathcal{S}}
\ket{\mathcal{R}, \mathcal{S}}^M, ~~~~
\label{eq:marshallsign}
\eea
where all coefficients are strictly positive, i.e.,
$c_{\mathcal{R}, \mathcal{S}}>0$;
the sign $(-)^\Gamma$ is defined by
$\Gamma={\sum_{1\le c_j \le L_x ,1\le i \le N_{c_j}}
(\frac{1}{2}-\beta^{c_j}_i)}$.
The total spin of $\ket{\Psi_G^M}$ is $S=|M|$ for
$|M|> \frac{1}{2}\Delta N$, and $S=\Delta N /2$ for
$\Delta N /2 \le M\le \Delta N /2$, respectively, where
$\Delta N$ is the difference between total particle
numbers in the $p_x$- and $p_y$-orbitals.
}

Theorem 3 can be proved following the proof of the Lieb-Mattis Theorem \cite{lieb1962a}
and of Lieb's Theorem \cite{lieb1989} for antiferromagnetic
Heisenberg models in bipartite lattices.
Here  $p_x$- and $p_y$-orbitals play the role of two sublattices.
However, the system here is itinerant not of local spin moments.
Because of the quasi-1D geometry, fermions do not pass each other,
and thus their magnetic properties are not affected by the mobile
fermions.
The detailed proof is presented in Appendix \ref{sect:appd_theorem2}.
However, this theorem cannot be generalized to the 3D case and the SU($N$)
case, even in 2D, because in both cases the antiferromagnetic coupling
$J<0$ leads to intrinsic frustrations.

The search for FM states has become a research focus in cold atoms
\cite{duine2005,zhangSZ2010,berdnikov2009,conduit2009,leblanc2009,
jo2009,pekker2011,cui2013}.
Both the 2D and 3D Hamiltonians $H_{kin}+H_{int}$ can be realized in the
$p$-orbital band in optical lattices.
With a moderate optical potential depth $V_0/E_R=15$ where $E_R$ is the
recoil energy, it was calculated that $t_\perp/t_\pp\approx 5\%$
\cite{isacsson2005}, and thus the neglect of $t_\pp$ in Eq. (\ref{eq:H_band})
is justified.
A Gutzwiller variational approach has been applied to the 2D Hamiltonian
of $H_{kin}+H_{int}$ \cite{wang2008}.
Furthermore, many transition-metal oxides possess $t_{2g}$-orbital
bands with quasi-2D layered structures, such as the ($001$) interface
of $3d$-orbital transition-metal oxides\cite{lilu2011,bert2011,chen2013}.
Its $3d_{xz}$ and $3d_{yz}$-bands are quasi-1D as described by
Eq. (\ref{eq:H_band}) with $p_{x(y)}$ there corresponding to $d_{x(y)z}$.
Also, strongly correlated $3d$ electrons possess the large $U$ physics.
Further discussion on the physics of finite $U$ and $V$ is given in Appendix \ref{app:finitUV}

\textit{Summary.--} We have shown - contrary to the normal situation in 1D without orbital degrees of freedom -
that fully saturated ferromagnetism is possible in certain tight-binding lattice
models with several orbitals at each site.
This holds for 2D and 3D models and for $SU(N)$ models as well as $SU(2)$ models.
Hard-core interactions in 1D chains, together with Hund's rule coupling,
stabilize the effect and result in unique ground states with saturated ferromagnetism.
The result also holds for a large region of electron densities in both
2D and 3D,
or in 1D with 2 or 3 p-orbitals at each site.
Our theorems might provide a reference point for the study of itinerant
FM in experimental orbitally active systems with
ultra-cold optical lattices and transition-metal oxides.

\textit{Acknowledgments.--}
YL and CW are supported by grants NSF DMR-1105945 and AFOSR
FA9550-11-1-0067(YIP).
EL is supported by NSF grants PHY-0965859 and  PHY-1265118.
YL thanks the Inamori Fellowship and the support at the Princeton Center
for Theoretical Science.
CW acknowledges the support from the NSF of China under Grant No. 11328403
and the hospitality of Aspen Center of Physics.
We thank S. Kivelson for helpful discussions and encouragement during
this project and we thank D. C. Mattis and H. Tasaki for helpful
comments on a draft of this paper.

\appendix

\section{Expressions for $U, V, J$ and $\Delta$}
\label{sect:int_para}

In this appendix, we present the expression for the interaction matrix
elements $U$, $V$, $J$ and $\Delta$ in $H_{int}$ defined in Eq. (\ref{eq:H_int}) in the body text.
We assume that the bare interaction between two particles in free space
is $V(\mathbf{r}_1-\mathbf{r}_2)$.
For example, it can be the Coulomb interaction between
electrons, or a short-range $s$-wave scattering interaction
between two ultra-cold fermion atoms.
Let us consider one site with degenerate $p_x$ and $p_y$ orbitals
whose Wannier orbital wave functions are $\phi_x(\mathbf{r})$ and
$\phi_y(\mathbf{r})$, respectively.
Then $U$, $V$, $J$ and $\Delta$ can be represented \cite{hubbard1963, slater1963} as
\bea
U&=&\int \textrm{d} \mathbf{r}_1  \textrm{d} \mathbf{r}_2  \phi_x (\mathbf{r}_1) \phi_x (\mathbf{r}_2)
V(\mathbf{r}_1-\mathbf{r}_2) \phi_x (\mathbf{r}_2) \phi_x (\mathbf{r}_1), \nn \\
V&=& \int \textrm{d} \mathbf{r}_1 \textrm{d}\mathbf{r}_2 \phi_x (\mathbf{r}_1) \phi_y (\mathbf{r}_2)
V(\mathbf{r}_1-\mathbf{r}_2) \nn \\
&&\times
\Big\{
\phi_y (\mathbf{r}_2) \phi_x (\mathbf{r}_1) -
\phi_x (\mathbf{r}_2) \phi_y (\mathbf{r}_1) \Big\},
\nn \\
J&=&2\int \textrm{d} \mathbf{r}_1 \textrm{d}\mathbf{r}_2 \phi_x (\mathbf{r}_1) \phi_y (\mathbf{r}_2)
V(\mathbf{r}_1-\mathbf{r}_2) \phi_x (\mathbf{r}_2) \phi_y (\mathbf{r}_1), \nn \\
\Delta &=& \int \textrm{d} \mathbf{r}_1 \textrm{d}\mathbf{r}_2
\phi_x (\mathbf{r}_1) \phi_x (\mathbf{r}_2) V(\mathbf{r}_1-\mathbf{r}_2)
\phi_y (\mathbf{r}_2) \phi_y (\mathbf{r}_1). \nn 
\eea

The physical meanings of $U, V, J$ and $\Delta$ can be
explained as follows.
Consider a single site with two orbitals
and put two fermions on the site.
There are four states in which each orbital is singly occupied,
including the triplet states
$p^\dagger_{x\uparrow} p^\dagger_{y\uparrow}|0 \rangle$,
$\frac{1}{\sqrt 2} \left\{ p^\dagger_{x\uparrow} p^\dagger_{y\downarrow}
+p^\dagger_{x\downarrow} p^\dagger_{y\uparrow}\right \} |0 \rangle$,
and $p^\dagger_{x\downarrow} p^\dagger_{y\downarrow}|0 \rangle$,
and the singlet state
$\frac{1}{\sqrt 2} \left \{p_{x\uparrow}^\dagger p_{y\downarrow}^\dagger
-p_{x\downarrow}^\dagger p_{y\uparrow}^\dagger \right\}\ket{0}$
with energies $V$ and $J+V$, respectively.
Their energy difference is the Hund's rule coupling energy.
The other two states are singlets involving doubly occupied orbitals, namely
$\frac{1}{\sqrt 2} \left \{p_{x\uparrow}^\dagger p_{x\downarrow}^\dagger\pm
p_{y\uparrow}^\dagger p_{y\downarrow}^\dagger \right\} \ket{0}$, whose
energies are $U\pm \Delta$, respectively.

\section{Proofs of Lemmas 2 and 3}
\label{app:lemmas}
In this appendix, we present the detailed proofs to Lemmas 2 and
3 which are used in proving Theorem 1.
Lemma 1, as we noted, is obvious.


\subsection {Proof of Lemma 2}
Let us start with the general basis $\ket{\mathcal{R}, \mathcal{S}}$ defined
in Eq. (\ref{eq:basis}) in the body text,
and
check the hopping matrix elements.
It suffices to consider hoppings along
the $x$-direction, because the $y$-direction is similar.
The following hopping along row $r$, denoted as
\bea
H_{x,\pm}(\mathbf{r}_i;\alpha^r_i)=-t_\pp p^\dagger_{x,\alpha^r_i}(\mathbf{r}_i\pm
\hat e_x) p_{x,\alpha^r_i} (\mathbf{r}_i),
\eea
generate non-zero off-diagonal matrix elements if $x^r_i+1 < x^r_{i+1}$,
or, $x^r_i-1> x^r_{i-1}$,
where the boundary condition ($\ast$) for coordinates is assumed
and the particle indices $i\pm 1$ are defined on row $r$ modulo $N_{r}$.
Without loss of generality, we only need to consider $H_{x,+}$.
If this hopping is not between the ends of the row,
when $H_{x,+}$ acts on $\ket{\mathcal{R}, \mathcal{S}}
=\ket{\{ \mathbf{r}^{r_j}_i \alpha^{r_j}_i; \mathbf{r}^{c_j}_i \beta^{c_j}_i\}}$,
it just replaces $p^\dagger_{x,\alpha^r_i}(\mathbf{r}_i)$ by
$p^\dagger_{x,\alpha^r_i}(\mathbf{r}_i +\hat e_x)$
in the sequence of creation operators in Eq. (\ref{eq:basis}) in the body text
without affecting the ordering;
thus its matrix element is just $-t_\pp$.
If this hopping goes from one end to another end, i.e., $x^r_i=L_x$, and then
$x^r_i+1 \equiv 1 \mod (L_x)$, it replaces the operator  $p^\dagger_{x,\alpha^r_i}
(\mathbf{r}_i)$ with $\mathbf{r}_i=(L_x,r)$ by that with $\mathbf{r}_i=(1,r)$
together with a minus sign if $N_{r}$ is even.
To fit the ordering of creation operators in Eq. (6) in the body text, 
we move the operator $p^\dagger_{x,\alpha^r_i}(\mathbf{r}_i)$ with
$\mathbf{r}_i=(1,a)$  to its right location
after passing $N_{r}-1$ operators in the $a$-th row.
If $N_{r}$ is even or odd, no additional sign is generated and the matrix
element is still $-t_\pp$.
The same reasoning applies to the hopping operator $H_{x,-}$, and
for those along the $y$-direction.

Next we check matrix elements associated with the interaction
terms in Eq. (\ref{eq:H_int}) in the body text.
On the physical Hilbert space ${\cal H}_{\mathcal{N}_{\mathcal{X}},
\mathcal{N}_{\mathcal{Y}}}$,
only the following term, denoted as
\bea
H_{J}(\mathbf{r})&=&-\frac{J}{2} \Big\{ p^\dagger_{x\uparrow}(\mathbf{r})
p_{x\downarrow}(\mathbf{r}) p^\dagger_{y\downarrow}(\mathbf{r}) p_{y\uparrow}(\mathbf{r})
+ h.c. \Big\}, \ \ \,
\eea
generates non-zero off-diagonal matrix elements.
When $H_{J}(\mathbf{r})$ acts on $\ket{\mathcal{R}, \mathcal{S}}
=\ket{\{ \mathbf{r}^{r_j}_i \alpha^{r_j}_i; \mathbf{r}^{c_j}_i \beta^{c_j}_i\}}$,
it updates the creation operators without affecting the ordering
in  Eq. (\ref{eq:basis}) in the body text, 
and thus the corresponding matrix elements
are just $-J/2$.
In summary, all the off-diagonal matrix elements are either zero or
negative, i.e., non-positive. \hfill \textit{Q.E.D.}

\subsection{Proof of Lemma 3}

We denote two general basis vectors $\ket {u}$ and $\ket {u^\prime}$
in ${\cal H}_{\mathcal{N}_{\mathcal{X}}, \mathcal{N}_{\mathcal{Y}}}^M$
as $\ket{u}=\ket{\mathcal{R}, \mathcal{S}}^M
=\ket{\{ \mathbf{r}^{r_j}_i \alpha^{r_j}_i; \mathbf{r}^{c_j}_i \beta^{c_j}_i\}}^M$
and $\ket{u^\prime} =\ket{\mathcal{R}', \mathcal{S}'}^M
=\ket{\{ \mathbf{r'}^{r_j}_i \alpha'^{r_j}_i; \mathbf{r'}^{c_j}_i \beta'^{c_j}_i\}}^M$.
First, we can successively apply the hopping terms to rearrange the
spatial locations of particles from $\mathcal{R}$ in $\ket{u}$ to be $\mathcal{R}'$.
We arrive at an intermediate
state $\ket{v}=\ket{\mathcal{R}', \mathcal{S}}^M
=\ket{\{ \mathbf{r'}^{r_j}_i \alpha^{r_j}_i; \mathbf{r'}^{c_j}_i \beta^{c_j}_i\}}^M$ as
\bea
\ket{v}
= \prod_{c_j,i} p^\dagger_{y,\beta^{c_j}_i} ( \mathbf{r'}^{c_j}_i )
\prod_{r_j,i} p^\dagger_{x,\alpha^{r_j}_i} (\mathbf{r'}^{r_j}_i) \ket{0}.
\label{eq:config_v}
\eea
Compared to the final state
$\ket{u^\prime}=\ket{\{ \mathbf{r'}^{r_j}_i \alpha'^{r_j}_i; \mathbf{r'}^{c_j}_i \beta'^{c_j}_i\}}^M$,
defined as
\bea
\ket{u^\prime}
= \prod_{c_j,i} p^\dagger_{y,\beta'^{c_j}_i} ( \mathbf{r'}^{c_j}_i )
\prod_{r_j,i} p^\dagger_{x,\alpha'^{r_j}_i} (\mathbf{r'}^{r_j}_i) \ket{0},
\label{eq:config_u}
\eea
the locations of particles in $\ket{v}$ and in $\ket{u^\prime}$
are equal, but the spin configuration in $\ket{v}$
are the same as that in $\ket{u}$.
This arrangement can be decomposed into independent hops within each
chain without interference among chains,
because particle numbers in each row and each column are
conserved separately.

Next we prove that it is possible to adjust the sequence of spin indices
in the chain of creation operators in Eq. (\ref{eq:config_v}) for $\ket{v}$
to be the same as that in Eq. (\ref{eq:config_u}) for $\ket{u^\prime}$.
Two sequences of spin indices are the same up to
a permutation.
Since any permutation can be decomposed into a product of
exchanges, we only need to prove that any exchange can be realized
by successively applying off-diagonal Hamiltonian matrix elements
of the hopping and $J$ terms.
Obviously, we only need consider the exchange of two opposite spins.

\begin{figure}[tbp]
\centering\epsfig{file=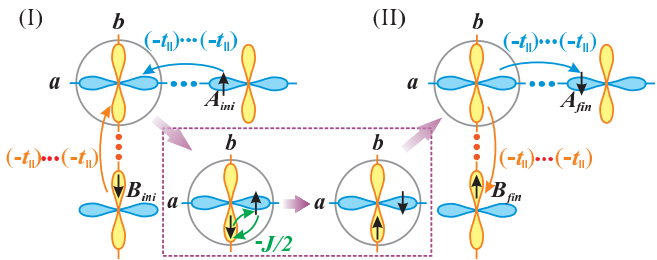,clip=1,width=\linewidth,angle=0}
\caption{The scheme of exchanging any two given opposite spins
in orthogonal chains.
Starting from configuration (I), two particles, marked with black
arrows for spin up and down, 
arrive at the crossing site (as circled) by successively hopping
along each chain.
Then their spins are flipped by the on-site $J$ term.
Finally, they hop back to the initial locations with spins
configuration flipped as in (II).
}
\label{fig:transitive}
\end{figure}

First, we consider the exchange between two particles $A$ and $B$
in orthogonal chains.
Without loss of generality, we assume $A$ to be in the $p_x$-orbital
of the $a$-th row and $B$ to be in the $p_y$-orbital of the $b$-th column
as shown in Figure \ref{fig:transitive}.
Their configuration is denoted as $A_{ini}$:$(\mathbf{r}_A;p_x\uparrow)$ and
$B_{ini}$:$(\mathbf{r}_B;p_y\downarrow)$ with $\mathbf{r}_A=(x,a)$ and
$\mathbf{r}_B=(b,y)$.
Since there is at least one hole in each chain, cyclic permutations
of particle locations along the chain can be realized by applying
only hopping terms along it.
We move these two particles to the crossing site $\mathbf{r}_c=(b,a)$ and
flip their spins by using the $J$ term.
We can then restore the spatial locations of particles in the $a$-th row
and the $b$-th column to be the same as those in $\ket{v}$ by applying only
hopping terms.
The net effect is the exchange of spin indices into
$A_{fin}$:$(\mathbf{r}_A;p_x\downarrow)$ and $B_{fin}$:$(\mathbf{r}_B;p_y\uparrow)$.

Second, we consider the exchange between two particles with opposite
spin indices in the same chain, or, in two parallel chains.
Without loss of generality, they may be assumed to be
in the $p_x$-orbitals in row $a_1$ and $a_2$ respectively.
Their coordinates and spins are denoted as
$A_{ini}(\mathbf{r}_A;p_x\uparrow)$ and $B_{ini}(\mathbf{r}_B;p_x\downarrow)$
with $\mathbf{r}_A=(m,a_1)$ and $\mathbf{r}_B=(n,a_2)$, respectively.
Let us choose an arbitrary $p_y$ particle and, without loss of generality,
assume its configuration to be $C(\mathbf{r}_C;p_y\downarrow)$
with $\mathbf{r}_C=(b,y)$.
Then we first exchange particles $A$ and $C$ following the method
described above, and then exchange particles $B$ with the updated
configuration of $C$.
The net effect is the exchange between $A_{ini}$ and $B_{ini}$ with the new
configuration of $A_{fin}(\mathbf{r}_A;p_x\downarrow)$ and $B_{fin}(\mathbf{r}_B;p_x\uparrow)$,
while $C$ is restored to its initial configuration.
Thus we have proved the transitivity of the Hamiltonian
matrix in the subspace ${\cal H}^M_{\mathcal{N}_{\mathcal{X}}, \mathcal{N}_{\mathcal{Y}}}$.
 \hfill \textit{Q.E.D.}

\subsection{More extensions}
In fact, Theorem 1 can be made even more general by adding
off-site interactions such as
\bea
H_{int}^\prime &=&\sum_{\mathbf{r}\mathbf{r}';\mu \nu}
\left( V_{\mathbf{r}\mathbf{r}';\mu \nu}
n_{\mu}(\mathbf{r}) n_{\nu}(\mathbf{r}')
-
J_{\mathbf{r}\mathbf{r}';\mu \nu}
\vec S_{\mu}(\mathbf{r}) \cdot \vec S_{\nu} (\mathbf{r}')\right), \nn \\
\eea
where $\mu, \nu$ represent orbital indices.
In order to satisfy the hypothesis of the Perron-Frobenius theorem, the spin
channel interaction parameters should be ferromagnetic, i.e.,
$J_{\mathbf{r}\mathbf{r}';\mu \nu }>0$, while the charge channel interactions
$V_{\mathbf{r}\mathbf{r}';\mu \nu}$ can be arbitrary.

\subsection{Discussion of Lemma 3 of transitivity}
\label{app:transitivy}
If the transitivity condition of the Hamiltonian matrix is not
satisfied, then Theorem 1 may not be valid, \textit{i.e.}, the ground
state might be degenerate.
We consider below a concrete example in which
all the rows of $p_x$-orbitals are
{\it empty} except in the first row where {\it all} the $p_x$-orbitals
are filled.
Thus particles in the first row cannot hop.
For the first row, all the different spin configurations are degenerate
because of the absence of hopping.
Let us assume that all other columns contain at least one hole.
Following Hund's rule, for every column of the $p_y$-orbital, say,
the $r$-th one, we align all
the particles therein to be the same as the one in the $p_x$-orbital at
site $(r,1)$.
Although the total spin for each column is fully polarized,
no coupling exists between adjacent columns, and thus the 2D system
overall is still paramagnetic.
Nevertheless, if we just add one particle in the 2nd row of the
$p_x$-orbital which is otherwise empty, it connects different columns
through multiple spin-flip processes from the $J$ term, and
realizes the transitivity condition.
The ground state is again unique and fully-polarized.

\textit{Condition} ($\ast \ast$) is sufficient but not necessary for Lemma 3 of
transitivity.
It would be interesting to figure out the necessary condition.
In fact, condition ($\ast \ast$) can be further weakened as follows:
There is at least one hole in \emph{one} of the chains along any one direction and
one hole in \emph{each} chain along other directions.
At the same time, there must be at least one particle in \emph{one}
of the columns and another particle in \emph{one} of the rows.

In particular, the situation is more complicated for the open boundary condition.
Although Lemma 2 of non-positivity is valid regardless of the oddness
of filling numbers in every chain,
it is more difficult to effect the connectivity with open
boundary conditions.
Nevertheless, we expect that in the thermodynamic limit
the effects of boundary conditions vanish, and the ground
state ferromagnetism remains robust for generic fillings.

\section{The Perron-Frobenius Theorem and Transitivity}
\label{app:p-f}

To keep the paper self-contained, we explain how
transitivity gives rise to a unique ground state
in the Perron-Frobenius set up \cite{perron1907, frobenius1908}.
Suppose $M$ is a real symmetric matrix with all off-diagonal elements non-positive.
Let $V$ be a ground state. Then, by the variational principle,
$|V|=\{ |V_j|\}$ is also a ground state.
If the ground state is unique, then $V=|V|$, \textit{i.e.}, $V_j \geq 0$ for all $j$.

Suppose now that $W$ is another ground state. Clearly, there is a real number $\alpha$
so that the ground state $\widetilde{V}=V+\alpha W$ has at lease one component,
say $\widetilde{V}_1$, equals zero.
Then $\hat V = |\widetilde{V}|$ is a ground state with non-negative components and
at least one component zero, namely $\hat V_1$.
Without loss of generality we may assume that the ground state eigenvalue
$\lambda$ is not zero and the diagonal elements $M_{ii}$'s are all negative,
for otherwise, we can replace $M$ by $M-cI$.
We thus have, for $p \in \mathbb{N}$, $M^p \hat V = \lambda^p \hat V \neq 0$,
but $(M^p \hat V)_1=0$.

Assuming transitivity now, we have that for some $p$, $(M^p)_{1j}$ has a strictly
non-zero entry for some $j$ such that $\hat V _j \neq 0$. This contradicts the fact
that $(M^p \hat V)_1=0$.

Thus, transitivity implies that every ground state has only non-zero components.
This means that there is no other ground state $W$, for otherwise the ground state
$(V + \alpha W)_j =0$ for some $\alpha$ and some $j$.

\section{Extension of Theorem 1 to SU($N$) symmetric systems}
\label{sect:SUN}


In this appendix, we extend Theorem 1 from the SU(2) systems
to those with SU($N$) symmetry.

The physical meanings of the $U$, $V$, $J$ and $\Delta$ in the
SU($N$) \textit{multi-orbital} interaction defined in Eq. (\ref{eq:H_suint}) in the body text
are similar to the case of $SU(2)$.
Again for simplicity, we consider the 2D case with $p_x$ and $p_y$ orbitals.
If we load two fermions in a single site, there are
${{2N} \choose 2}=N(2N-1)$ states
which are SU($N$) rank-2 tensor states.
They can be classified into a) one set of symmetric tensor states,
b) one set of anti-symmetric tensor states with singly occupied orbitals,
c) two sets of anti-symmetric tensor states with doubly occupied orbitals.
Their energies are $V$, $V+J$ and $U\pm \Delta$, respectively.
The dimensions for the rank-2 SU($N$) symmetric and anti-symmetric tensor
representations are $N(N\pm 1)/2$, respectively.

Again Lemma 1 for the SU(2) case remains valid for the SU($N$) Hamiltonian
in the limit  $U \rightarrow +\infty$.
The many-body basis for the SU($N$) case can still be set up in a manner
 similar to that defined in Eq. (\ref{eq:basis}) in the body text.
The only difference is that fermion spins can take $N$ different values.
The off-diagonal elements of the Hamiltonian matrix
in ${\cal H}_{N_A, N_B}$ are also non-positive, and thus Lemma 2
remains valid.
Following essentially the same method as in the proof of Lemma 3
with slight variations,
any two bases in the subspace ${\cal H}_{\mathcal{N}_{\mathcal{X}},
\mathcal{N}_{\mathcal{Y}}}^{\mathcal{N}_{\sigma}}$
can be connected by successively applying the hopping and $J$ terms
under  \textit{condition} ($\ast \ast$).
Thus the Hamiltonian matrix is also transitive in
each subspace ${\cal H}_{\mathcal{N}_{\mathcal{X}}, \mathcal{N}_{\mathcal{Y}}}
^{\mathcal{N}_{\sigma}}$.

The SU(2) fully polarized FM state with total spin $S=N_{tot}/2$
can be easily generalized to the SU($N$) case.
These SU($N$) FM states belong to the representation denoted by
the Young pattern with one row of $N_{tot}$ boxes, {\it i.e.},
the fully symmetric rank-$N_{tot}$ tensor representation.
Its dimension,
${ {N+N_{tot}-1} \choose {N_{tot}} }=\frac{(N+N_{tot}-1)!}{(N-1)!N_{tot}!}$
, is the number
of partition of $N_{tot}$ particles into $N$ different components, which is
just the number of different subspaces ${\cal H}_{\mathcal{N}_{\mathcal{X}},
\mathcal{N}_{\mathcal{Y}}}^{\mathcal{N}_{\sigma}}$
with respect to the configurations of ${\mathcal{N}_{\sigma}}$.
Any state of this representation is fully symmetric with respect to
exchange spin components of any two particles.

Since Lemmas 1, 2, and 3 are generalized to the SU($N$) case,
we obtain Theorem 3.
\hfill \textit{Q.E.D.}

\section{FM in the 3D cubic lattice}
\label{sect:cubic}

In this appendix, we generalize Theorem 1 to the 3D Hamiltonian
$H_{kin}+H_{int}$ in the same limit $U \rightarrow
\infty$ with $J>0$.

The generalization is easy. The particle number in each
chain along any of the three directions is separately conserved
because of the vanishing of transverse hoppings and the absence
of doubly occupied orbitals.
We can further set up the many-body basis in a manner similar to Eq. (\ref{eq:basis}) in the body text
by ordering particles in each chain and ordering
one chain after another.
The non-positivity of the off-diagonal elements of the many-body Hamiltonian
matrix is still valid under \textit{condition}($\ast$).
Next, we generalize Lemma 3 of transitivity to 3D.

{\lemma \textbf{(Transitivity of the 3D Hamiltonian)} The many-body Hamiltonian matrix
of the 3D version of $H_{kin}+H_{int}$ is transitive under  \textit{condition} ($\ast \ast$)
in the Hilbert subspace characterized by the particle number
distributions in each chain and the $z$-component of total spin.
}

{\proof} The proof is very similar to that of Lemma 3. We only need to
show that spin configurations of any two particles $A$ and $B$, if different,
can be exchanged by applying hopping and $J$ terms.
Lemma 3 has already proved that it is true if the two particles are coplanar.
Now we consider the non-coplanar case, and denote particle locations
as $\mathbf{r}_A$ and $\mathbf{r}_B$, respectively.
If they lie in parallel orbitals, say, $p_x$-orbital, we can find
an $x$-directional chain with its $yz$ coordinates $(r_{A,y},r_{B,z})$;
if they lie in orthogonal orbitals, say,
particle $A$ lying in the $p_x$-orbital and particle $B$ lying
in the $p_y$-orbital, we can find a $z$-directional chain with the
$xy$ coordinates $(r_{B,x}, r_{A,y})$.
In both cases, the third chain defined above is coplanar with
each of the two particles $A$ and $B$.

We then choose a particle $C$ in the third chain.
Let us consider the general SU($N$) case.
If the spin component of particle $C$ is the same as one of
the two particles, say, particle $B$, owing to
Lemma 3, we can first switch the spin configuration between
$A$ and $C$, and then that between $B$ and $C$.
If the spin component of particle $C$ is different from both
that of $A$ and $B$, we first switch the spin configuration
between $A$ and $C$, then that between $B$ and $C$, and at
last that between $A$ and $C$.
The net result is that the spin configuration between $A$ and $B$
is switched while that of $C$ is unchanged.
\hfill \textit{Q.E.D.}

Since all the three lemmas have been generalized to the 3D case,
we arrive at Corollary 1 of ferromagnetism in 3D in the main
text. \hfill \textit{Q.E.D.}

\section{FM in the 1D lattice}
\label{sect:1dFM}

\begin{figure}[tbp]
\centering\epsfig{file=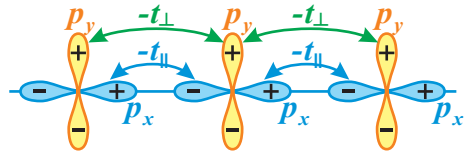,clip=1,width=0.7\linewidth,angle=0}
\caption{The 1D lattice along $x$ direction with $p_x$- and $p_y$-orbitals at
each site.
Different from Fig. 1 in the main text, 
 here, particles in $p_x$- and $p_y$-orbitals
can all move along the $x$-direction with hopping amplitudes $t_{\pp}$ and $t_{\perp}$
respectively.
The signs of $t_{\pp}$ and $t_{\perp}$ can be changed independently by gauge
transformations.
}
\label{fig:1dorbital}
\end{figure}

As a byproduct, our results can be extended to 1D multi-orbital systems.
As illustrated in Fig. \ref{fig:1dorbital},
in addition to the $\sigma$-bonding with hopping amplitude $t_\pp$, a nonzero
$\pi$-bonding with hopping amplitude $t_\perp$ is needed in the kinetic Hamiltonian,
 to satisfy Lemma 3.
Unique FM ground states in this 1D system can be proved under the same \textit{conditions}
($\ast$) and ($\ast \ast$) as the following corollary. We emphasize that this result was
already obtained by Shen \cite{shen1998} using the Bethe Ansatz.

\vspace{-2mm}
{\corollary \textbf{(1D FM Ground State)}
The statements in Theorems 1 and 2 of FM are also valid for the
1D multi-orbital systems $H'_{kin}+H_{int}$ under the same conditions.
Here,
\bea
&&H'_{kin}= \sum_{x=1, \sigma=\uparrow,\downarrow}^{L_x}
\big[-t_{\perp} \sum_{\mu=y,(z)} p_{\mu,\sigma}^\dagger(x+1)p_{\mu,\sigma}(x) \nn \\
&&
\, \, \, \, \, -t_{\pp} p_{x,\sigma}^\dagger(x
+1)p_{x,\sigma}(x)+ h.c.\big]
-\mu_0 \sum_{x=1}^{L_x} n(x).
\eea
}

\section{Proof of Theorem 3 on the  absence of FM}
\label{sect:appd_theorem2}

In this appendix, we will consider the opposite situation with $J<0$,
{\it i.e.}, anti-Hund's rule coupling.

\subsection{FM states as the highest energy states}

A direct result of the anti-Hund's rule coupling is the following corollary.
{\corollary
Consider the same Hamiltonian in the same limits as those
in Theorem 1 but in the case of $J<0$.
Under \textit{condition} ($\ast$), the many-body eigenstates with the highest energy
include the fully polarized states.
If  \textit{condition} ($\ast \ast$) is also satisfied, the highest energy states are
non-degenerate except for the trivial spin degeneracy.}

{\proof As we discussed before, the sign of the hopping integral
$t_\pp$ can be flipped by the gauge transformation
$p_{\mu,\sigma}(\mathbf{r})\rightarrow (-)^{r_\mu}  p_{\mu,\sigma}(\mathbf{r})$.
We denote the resultant Hamiltonian as $H^\prime$,
whose eigenstates have the same energy and the same physical properties
as those of $H$.
The negative of $H^\prime$, {\it i.e.}, $-H^\prime$, satisfies all conditions
of Theorem 1.
The ground states of $-H^\prime$ are the highest energy states
of $H^\prime$, and thus correspond to the highest energy states of $H$
up to a gauge transformation, which proves this corollary. \hfill
\textit{Q.E.D.}}

\subsection{Proof of Theorem 3}
Following the same strategy in the proof of Lieb-Mattis' Theorem \cite{lieb1962a}
and Lieb's Theorem \cite{lieb1989} for antiferromagnetic
Heisenberg models on bipartite lattices,
we first perform a gauge transformation on the operators
for $p_y$-orbitals and keep those of $p_x$-orbitals unchanged
\bea
p^\prime_{x,\alpha}(\mathbf{r})=p_{x,\alpha}(\mathbf{r}), \ \ \,
p^\prime_{y,\alpha}(\mathbf{r})=
(-)^{\alpha-\frac{1}{2}} p_{y,\alpha}(\mathbf{r}).
\label{eq:gauge_2}
\eea
After this transformation, $H$ is transformed to $H^\prime$, which
is identical to $H$ except that the $xy$-components of
the Hund's coupling term flip the sign as
\bea
H_{J}^\prime&=& -|J| \sum_{\mathbf{r}} \Big\{ S_x^x(\mathbf{r}) S_y^x(\mathbf{r})
+S_x^y(\mathbf{r}) S_y^y (\mathbf{r})
-S_x^z(\mathbf{r}) S^z_y(\mathbf{r}) \Big\}, \nn \\
\eea
and the many-body bases defined in Eq. (\ref{eq:basis}) in the body text
transform
as
\bea
\ket{\mathcal{R}, \mathcal{S}}^\prime=
(-)^\Gamma
\ket{\mathcal{R}, \mathcal{S}},
\eea
with
\bea
\Gamma={\sum_{1\le c_j \le L_x ,1\le i \le N_{c_j}} (\frac{1}{2}-\beta^{c_j}_i)}.
\eea
For each subspace ${\cal H}^M_{\mathcal{N}_{\mathcal{X}}, \mathcal{N}_{\mathcal{Y}}}$,
the matrix element of $H^\prime$ satisfies Lemma 2 of non-positivity under
\textit{condition} ($\ast$), and Lemma 3 of transitivity under
\textit{condition} ($\ast \ast$).
Again, the Perro-Frobenius theorem ensures that the ground state
$\ket{\Psi_G^M}$ in each subspace  ${\cal H}^M_{\mathcal{N}_{\mathcal{X}},
\mathcal{N}_{\mathcal{Y}}}$
is non-degenerate, and
\bea
\ket{\Psi_G^M}
&=& \sum_{\mathcal{R}, \mathcal{S}}
  (-)^\Gamma 
c_{\mathcal{R}, \mathcal{S}} 
\ket{\mathcal{R}, \mathcal{S}} 
=\sum_{\mathcal{R}, \mathcal{S}}
 c_{\mathcal{R}, \mathcal{S}}
 \ket{\mathcal{R}, \mathcal{S}}^\prime
\eea
with $c_{\mathcal{R}, \mathcal{S}}>0$.

Next we study the spin quantum number for the state $\ket{\Psi^M_G}$.
Following the method in Ref. \cite{lieb1962},
we define a reference Hamiltonian,
\bea
H^R&=& |J| \Big\{ \sum_{\mathbf{r}} \vec S_x (\mathbf{r}) \Big\} \cdot
\Big\{ \sum_{\mathbf{r}} \vec S_y (\mathbf{r}) \Big\}.
\label{eq:H_ref}
\eea
The spectra of Eq. (\ref{eq:H_ref}) can be easily solved as
\bea
E(S_x,S_y;S)&=& |J| \big\{ S(S+1)-S_x(S_x+1) \nn \\
&&-S_y(S_y+1)\big \}/2,
\eea
where $S_x$ ($S_y$) is the total spin of all the particles in
the $p_x$ ($p_y$)-orbital, respectively; $S$ is the total
spin of the system which takes value from $|S_x-S_y|, |S_x-S_y|+1, \cdots,
S_x+S_y$.
For any fixed values of $S_x$ and $S_y$, the minimization of $E(S_x,S_y;S)$
is reached at $S=|S_x-S_y|$ which yields the result:
\bea
E_{min}(S_x,S_y)=-|J|\big\{S_x S_y + \mbox{min} (S_x,S_y)\big\}.
\eea

Define $N_x=\sum_{1\le r_i \le L_y}  N_{r_i}$ and $N_y=\sum_{1\le c_i \le L_x}
N_{c_i}$, and thus $S_x \le N_x/2$ and $S_y \le N_y/2$.
The absolute ground state energy for $H^R$ is reached with
\bea
S_x=N_x/2, \ \ \, S_y= N_y/2, \ \ \, S=\Delta N/2.
\eea
Thus in all the subspaces of ${\cal H}^M_{N_A, N_B}$ with
$M\le \Delta N/2$, the ground states of $H^R$,
$\ket{\Psi_G^{M,R}}$, possess the spin quantum number
$S=\Delta N/2$.
In comparison, for $M>\Delta N/2$, the spin quantum number
of $\ket{\Psi_G^{M,R}}$ is $S=M$.
Clearly, by the same transformation given in Eq. (\ref{eq:gauge_2}),
the $H^R$-matrix satisfies Lemma 2 in each subspace
${\cal H}^M_{\mathcal{N}_{\mathcal{X}}, \mathcal{N}_{\mathcal{Y}}}$.
Since the Lemma 3 of transitivity is not satisfied for $H^R$,
its ground states $\ket{\Psi^{M,R}_G}$ are expressed as
\bea
\ket{\Psi_G^{M,R}}
&=& \sum_{\mathcal{R}, \mathcal{S}}
  (-)^\Gamma
c^R_{\mathcal{R}, \mathcal{S}} 
\ket{\mathcal{R}, \mathcal{S}}^M
\eea
with $c^R_{\mathcal{R}, \mathcal{S}}\ge 0$.
Nevertheless $\ket{\Psi^{M,R}_G}$ carries a unique spin quantum
number as analyzed above.

Now we are ready to prove Theorem 3.
Obviously $\avg{\Psi_G^{M,R}|\Psi_G^M}>0$, thus $\Psi_G^M$ shares
the same spin quantum number $S$ as that of $\ket{\Psi_G^{M,R}}$.
In short, $\ket{\Psi_G^M}$ is the non-degenerate ground state in the subspace
${\cal H}^M_{\mathcal{N}_{\mathcal{X}}, \mathcal{N}_{\mathcal{Y}}}$.
For the series of subspaces ${\cal H}^{M}_{\mathcal{N}_{\mathcal{X}},
\mathcal{N}_{\mathcal{Y}}}$
with $M \le \Delta N/2$, $\ket{\Psi_G^M}$'s form spin multiplets
with $S=\Delta N/2$.
Thus we conclude that the ground state energies $E_G^M$ in each subspace
${\cal H}^{M}_{\mathcal{N}_{\mathcal{X}}, \mathcal{N}_{\mathcal{Y}}}$ satisfy
$E_G^{M}< E_G^{M^\prime}$ for $\Delta N/2 < M<M^\prime$,
and $E_G^M=E_G^{\Delta N/2}$ for $M\le \Delta N/2$.
\hfill \textit{Q.E.D.}

\subsection{More extensions}

Theorem 3 implies strong ferromagnetic correlation inside and among parallel
chains.
Consider the special case in which there is only one particle in each column,
while in all the rows particle densities are positive in the thermodynamic
sense. Then the ground states are nearly fully polarized.
Even though the inter-orbital coupling $J$ is antiferromagnetic, the
particles in the column mediate FM coupling among those in the rows.
If particle numbers in rows and columns are equal, the ground state is
a spin singlet.
Although we cannot prove it, a spontaneous symmetry breaking
spin-nematic ground state conceivably occurs in the thermodynamic limit.
All the rows and columns are FM ordered, but the polarizations
of rows and columns are opposite to each other.
The possibility of spin-nematic phase also applies to the 3D version
of the SU(2) Hamiltonian $H_{kin}+H_{int}$.
In this case, similar to the frustration in the 2D triangular lattice,
the FM polarizations in the three types of orthogonal chains may form a
120$^\circ$ angle with respect to each other.

\section{Further discussion}
\label{app:finitUV}
In this appendix, we estimate the FM energy scale $J_{FM}$ and the effect
of finite values of $U$ which result in an antiferromagnetic (AFM) energy
scale $J_{AFM}$ (See Section \ref{app:AFM} below).

\subsection{The FM energy scale $J_{FM}$}
We assume that the electron filling in every chain is the same.
The average density per orbital (not per site) is $x$ which
satisfies $0<x<1$, and then the average distance between two
adjacent fermions in the same chain is $d=1/x$.
The FM energy scale $J_{FM}$ is estimated as the energy cost of
flipping the spin of one fermion while keeping all other fermions spin
polarized.
$J_{FM}$ determines the spin-wave stiffness and sets up the energy
scale of Curie temperature.
For simplicity, we only consider the 2D spin-$\frac{1}{2}$ case
as an example.

Let us first consider the \emph{low filling} limit $x\ll 1$, and
start with the fully spin polarized ground state as a background.
Without loss of generality, we choose the first row of
$p_x$-orbital, and pick up the $i$-th $p_x$-orbital fermion in this row.
We consider the motion of the $i$-th fermion while fixing positions
of all other fermions.
The locations of the $i\pm 1$-th fermions are the wavefunction nodes of the
$i$-th one, and the typical distance between the $i$-th and $i\pm 1$-th
fermions is $d$.
In fact, typically speaking, before the $i$-th fermion sees these nodes,
it feels the scattering potential of $V$ from two $p_y$-orbital fermions
intersecting this row with the average distance of $d$.
If we flip the spin of the $i$-th fermion, then the scattering
potential from its adjacent $p_y$-orbital fermions increases to the
order of $J+V$.
Under the condition that $xt/V\ll 1$, we can estimate from strong
coupling analysis that the energy cost of is the order of
$J_{FM}\sim x^3 \frac{t^2}{V} \frac{J}{J+V}$.
In the limit of $J\gg V$, $J_{FM}$ saturates to the order
of $x^3\frac{t^2}{V}$.

On the other hand, at the \emph{high filling} limit, {\it i.e.}, $1-x\ll 1$,
although on most sites two fermions are spin polarized by the
Hund's rule coupling $J$, the intersite FM coherence is mediated
by the motion of holes, and thus, the FM energy scale is much smaller than $J$.
In fact, in the absence of holes, i.e., $x=1$, all the spin configurations
are degenerate which suppresses $J_{FM}\rightarrow 0$.
The average distance between holes along the same chain is $d_h=1/(1-x)$.
Again let us start with a fully spin polarized background.
Without loss of generality, we pick up a spin-1 site at the
intersection of the $i$-th row and $j$-th column.
This site is filled by two fermions coupled by Hund's rule and
we flip its spin.
This process generates a new scattering center to adjacent
holes in the $i$-th row and in the $j$-th column, and the scattering
potential is at the order of $J$.
In case of $(1-x)t/J\ll 1$, this spin flipped site effectively
blocks the motion of holes, which costs kinetic energy at the order
of $J_{FM}\propto t(1-x)^2$.

Conceivably, $J_{FM}$ is optimized at certain intermediate filling $x$.
While generally evaluating $J_{FM}$ in this regime is
difficult, we can consider a special case of $x=1/2$, such that
the FM state coexists with the antiferro-orbital ordering.
The ideal N\'{e}el orbital configuration is that $p_x$ and $p_y$-orbitals
are alternatively occupied with spin polarized fermions.
In the case of $V\gg t$, the orbital superexchange is at the order
of $t^2/V$, and flipping the spin of one fermion reduces the orbital
superexchange energy to $t^2/(V+J)$.
The difference is the FM energy scale $J_{FM} \propto \frac{t^2}{V}
\frac{J}{J+V}$.

\subsection{The AFM energy scale $J_{AFM}$}
\label{app:AFM}
So far, we have only considered the case of infinite $U$ which
suppresses the AFM energy scale $J_{AFM}$ to zero.
At large but finite values of $U$, fermions in the same chain
with opposite spins can pass each other.
This process lowers the kinetic energy and sets up $J_{AFM}$.
In the low filling limit $x\ll 1$, the probability of two
fermions with opposite spins sitting on two neighboring sites
scales as $x^3$ under the condition that $x t/U \ll 1$,
and thus $J_{AFM} \sim x^3 \frac{t^2}{U}$.
At high fillings, $x\rightarrow 1$, the above probability simply
scales as $x$, and thus $J_{AFM} \sim x \frac{t^2}{U}$.

Let us compare the energy scales of $J_{FM}$ vs $J_{AFM}$.
At low filling limit, since usually $U\gg V$ and $J, V$ are
at the same order, $J_{FM}$ wins over $J_{AFM}$.
Nevertheless, $J_{AFM}$ increases with $x$ monotonically, and
thus it wins over $J_{FM}$ as $x\rightarrow 1$.
The FM ground states are expected to be stable in the
low and intermediate filling regimes until $J_{AFM}$
becomes comparable with $J_{FM}$.



\end{document}